\documentclass[conference]{IEEEtran}
\IEEEoverridecommandlockouts
\usepackage{cite}
\usepackage{amsmath,amssymb,amsfonts}
\usepackage{algorithmic}
\usepackage{graphicx}
\usepackage{textcomp}
\usepackage{xcolor}
\usepackage{subcaption}
\usepackage{hyperref}
\def\BibTeX{{\rm B\kern-.05em{\sc i\kern-.025em b}\kern-.08em
    T\kern-.1667em\lower.7ex\hbox{E}\kern-.125emX}}
\usepackage{xcolor, soul}
\begin{document}

\hyphenation{MagnaTag-ATune}

\def\x{{\mathbf x}}
\def\L{{\cal L}}
\def\choi{VGG-CNN}
\def\pons{MUSICNN}

\title{How Low Can You Go? Reducing Frequency and Time Resolution in Current CNN Architectures for Music Auto-tagging}

\author{\IEEEauthorblockN{Andres Ferraro, Dmitry Bogdanov, Xavier Serra}
\IEEEauthorblockA{\textit{Music Technology Group} \\
\textit{Universitat Pompeu Fabra}\\
Barcelona, Spain \\
andres.ferraro@upf.edu}
\and
\IEEEauthorblockN{Jay Ho Jeon, Jason Yoon}
\IEEEauthorblockA{\textit{Kakao Corp.}\\
Seoul, Republic of Korea \\
jason.yoon@kakaocorp.com }
}
\maketitle

\begin{abstract}
Automatic tagging of music is an important research topic in Music Information Retrieval 
and audio analysis algorithms proposed for this task have 
achieved improvements with advances in deep learning. In particular, many state-of-the-art systems use Convolutional Neural Networks and operate on mel-spectrogram representations of the audio. 
In this paper, we compare 
commonly used mel-spectrogram representations 
and evaluate model performances that can be achieved by reducing the input size in terms of both lesser amount of frequency bands and larger frame rates. 
We use the MagnaTagaTune dataset for comprehensive performance comparisons and then compare selected configurations on the larger Million Song Dataset. The results of this study can serve researchers and practitioners in their trade-off decision between accuracy of the models, data storage size and training and inference times.
\end{abstract}


\section{Introduction}\label{sec:introduction}

Current state-of-the-art systems for music auto-tagging using audio are based on deep learning, in particular convolutional neural networks (CNNs), following two different approaches, one directly using the audio as input (end-to-end models)~\cite{lee2017sample} and the other using the spectrograms as input~\cite{dieleman2014end,choi2017tutorial}. Previous works~\cite{pons2017end} suggest that two approaches can have a comparative performance when they are applied on large datasets.

We can distinguish two architectures for the spectrogram-based CNN solutions, depending on whether they use multiple convolutional layers of small filters~\cite{choi2017convolutional, choi2016automatic} or if they use 
multiple filter shapes~\cite{pons2017timbre,pons2017designing, pons2017end}. The former is borrowed from the computer vision field (VGG~\cite{simonyan2014very}) and gives a good performance without prior domain knowledge, while the latter is based on such a knowledge and employs filters designed to capture information relevant for music auto-tagging such as timbre or rhythm. Commonly mel-spectrograms are used with such architectures although constant-Q~\cite{oramas2018multimodal, choi2017tutorial} and raw short-time Fourier transform (STFT)~\cite{choi2018comparison} can be also applied.

In this paper, we compare the performance of two state-of-the-art CNN approaches commonly used for music auto-tagging~\cite{choi2016automatic,pons2017end}, using different mel-spectrogram representations as an input.
Our research question is whether it is possible for these approaches to operate on reduced data inputs without a significant drop in a prediction accuracy.

To this end, we consider 
reducing the size of the input spectrograms  in terms of both lesser amount of frequency bands and larger frame rates. 
We show that by reducing the frequency and time resolution we can train the network faster with a small decrease in the performance. The results of this study can help 
to build faster CNN models as well as reduce the amount of data to be stored and transferred optimizing resources when handling large collections of music.


\section{Related work}\label{sec:related_work}

In image processing, there are studies that consider simplifications of CNN architectures by means of reducing network width, depth and input resolution~\cite{tan2019efficientnet}. However, only few previous studies compared different spectrogram representations for CNN architectures in Music Information Retrieval (MIR). Instead, it is common to focus on tuning model hyper-parameters 
with a fixed chosen input. The choice of the spectrogram input is done empirically and often follows approaches previously reported in literature. 
Very few information comparing different inputs is available as the authors tend to only report the most successful approaches. 
Also, as the existing studies on music auto-tagging focus on optimizing accuracy metrics,  there is a lack of works that intend to simplify networks and their inputs for computational efficiency and 
consider practical aspects of the efficient ways to store spectrogram representations.

To the best of our knowledge, there is no systematic comparison of mel-spectrogram representations. The only work we are aware of in this direction has been done by Choi et al.~\cite{choi2018comparison}, where the authors compare model performances under different pre-processing strategies such as scaling, log-compression, and frequency weighting. The same authors provide an overview of different inputs that can be used for the auto-tagging task in~\cite{choi2017tutorial}. In relation to mel-spectrograms, they suggest that one can optimize the input to the network by changing some of the signal processing parameters such as sampling rate, window size, hop size or mel bins resolution. These optimizations can help to minimize data size and train the networks more efficiently, however, no quantitative evaluations are provided.





\section{Datasets}\label{sec:datasets}
Researchers in music auto-tagging commonly use the MagnaTagATune dataset~\cite{law2009evaluation} to evaluate multiple settings and then repeat some settings on Million Song Dataset~\cite{bertin2011million} to 
validate differences in performances on a larger scale~\cite{dieleman2014end,choi2016automatic,pons2017end}. It is important to note that both datasets contain unbalanced and noisy and/or weakly-labeled annotations~\cite{choi2017effects} and therefore are challenging to work with, as the reliability of conducted evaluations may be affected~\cite{sturm2012survey}. Still, these are the two mostly used datasets for benchmarking due to the availability of audio.

\subsection{MagnaTagATune (MTAT)}
MagnaTagATune 
dataset contains multi-label annotations of genre, mood and instrumentation for 25,877 audio segments. Each segment is 30 seconds long, and the dataset contains multiple segments per song. All the audio is in MP3 format with 32 Kbps bitrate and 16 KHz sample rate. The dataset is split into 16 folders, and researchers commonly use the first 12 folders for training, the 13th for validation and the last three for testing. Also, only 50 most frequent tags are typically used for evaluation. These tags 
include genre and instrumentation labels, as well as eras (e.g., '80s' and '90s') and moods.


\subsection{Million Song Dataset (MSD)}
The MSD~\cite{bertin2011million}  is a large dataset of audio features, expanded by the MIR community with additional information including tags, lyrics and other annotations. 
It also contains a subset mapped by researchers to 30 seconds audio previews available at 7digital and collaborative tags from Lastfm. This subset contains 241,904 annotated track fragments and is commonly used as another larger scale benchmark for music auto-tagging systems. The tags cover genre, instrumentation, moods and eras. Audio fragments vary in their quality, encoded as MP3 with a bitrate ranging from 64 to 128 Kbps and the sample rates of 22 KHz or 44 KHz. 


\section{Baseline architectures}\label{sec:baselines}
In this work, we reproduce two CNN architectures applying them on mel-spectrograms with reduced frequency and time resolution. These architectures are among the best performing according to the 
existing evaluations on the MTAT and MSD datasets: 


\begin{itemize}
\item
\textbf {VGG applied for music (\choi)}~\cite{choi2016automatic}.
This architecture contains multiple layers of small-size 2D-filters as it has been adapted from the computer vision field ~\cite{simonyan2014very}. It is a fully-convolutional network consisting of four convolutional layers with small 3$\times$3 filters\footnote{Number of mel bands $\times$ number of frames.} and max pooling (MP) settings presented in Table~\ref{tab:vgg}. The network operates on 96-bands mel-spectrograms for 29.1s audio segments, 12 KHz sample rate, 512 samples frame size and the hop size of 256 samples. 

\item 
\textbf{Musically-motivated CNN (\pons)}~\cite{pons2017end}. 
The architecture 
contains more filters of different shapes designed with an intention to capture musically relevant information such as timbre
(38$\times$1, 38$\times$3, 38$\times$7, 86$\times$1, 86$\times$3, 86$\times$7)
and temporal patterns (1$\times$32,  1$\times$64, 1$\times$128, 1$\times$165) in the first layer. The convolution results are concatenated and passed to three additional convolutional layers including residual connections.\footnote{We refer the readers to the original paper for all architecture details.} Original network operates on 96-bands mel-spectrograms computed on smaller 15s audio segments with 16 KHz sample rate, 512 samples frame size and 256 samples hop size.\footnote{Frame and hop size settings are confirmed in personal communication with the author.} It then averages tag activation scores across multiple segments of the same audio input.

\end{itemize}

For evaluation on MTAT and MSD, we use batch normalization,  Adam \cite{kingma2014adam} as optimization method with a learning rate of 0.001 and binary cross-entropy as loss function for both architectures following their authors. 


\begin{table}[!h]
 \caption{The baseline VGG CNN model architecture.}
 \begin{center}
 \begin{tabular}{|l|l|}
  \hline
\textbf{Input} & \textbf{Mel-spectrogram (96$\times$1366 $\times$ 1)} \\
\hline
Layer 1 & Conv 3$\times$3$\times$128 \\
& MP (2, 4) (output: 48$\times$341$\times$128) \\
Layer 2 & Conv 3$\times$3$\times$384 \\
& MP (4, 5) (output: 24$\times$85$\times$384) \\
Layer 3 & Conv 3$\times$3$\times$768 \\
& MP (3, 8) (output: 12$\times$21$\times$768) \\
Layer 4 & Conv 3$\times$3$\times$2048 \\
& MP (4, 8) (output: 1$\times$1$\times$2048) \\
\hline
Output & 50$\times$1 (sigmoid) \\
\hline
\end{tabular}
\end{center}

 \label{tab:vgg}
\end{table}

\section{Mel-spectrograms}\label{sec:melspectrograms}
We computed mel-spectrograms using typical setting for the MTAT dataset in the state of the art~\cite{choi2016automatic,pons2017end}. The most common settings are 12 KHz or 16 KHz sample rate, 
frame and hop size of 512 and 256 samples, respectively, and Hann window function. Commonly, 
96 or 128 mel bands are 
used, covering all frequency range below Nyquist (6 KHz and 8 KHz, respectively) and computed using Slaney's mel scale implementation~\cite{slaney1998auditory}.
To normalize the mel-spectrograms we considered two log-compression alternatives denominated as \textit{``dB''} for $
10 \cdot log_{10}(x)$ ~\cite{choi2018comparison} and \textit{``log''} for $
log(1+10000 \cdot x)$ ~\cite{dieleman2014end}.

\begin{table}[t]
\caption{
 Mel-spectrograms configurations evaluated on the MTAT dataset. Hop sizes are reported relative to the reference hop size of 256 samples (e.g., $\times$5 stands for a 5 times longer hop size).}
 \begin{center}
 \begin{tabular}{|c|c|c|c|c|c|}
\hline
\textbf{sample rate} & \textbf{\# mel} & \textbf{hop size} & \textbf{log type} \\
\hline
12 KHz & 128                    & $\times1, \times2, \times3, \times4 \times5, \times10$ & log, dB \\
12 KHz & 96                     & $\times1, \times2, \times3, \times4 \times5, \times10$ &log, dB \\
12 KHz & 48                     & $\times1, \times2, \times3, \times4 \times5, \times10$ &log, dB \\
12 KHz & 32                     & $\times1$&log, dB \\
12 KHz & 24                     & $\times1$&log, dB \\
12 KHz & 16                     & $\times1$&log, dB \\
12 KHz & 8                      & $\times1$&log, dB \\
16 KHz & 128                    & $\times1, \times2, \times3, \times4 \times5, \times10$ &log, dB \\
16 KHz & 96                     & $\times1, \times2, \times3, \times4 \times5, \times10$ &log, dB \\
16 KHz & 48                     & $\times1, \times2, \times3, \times4 \times5, \times10$ &log, dB \\
16 KHz & 32                     & $\times1$&log, dB \\
16 KHz & 24                     & $\times1$&log, dB \\
16 KHz & 16                     & $\times1$&log, dB \\
16 KHz & 8                      & $\times1$&log, dB \\
\hline
 \end{tabular}
\end{center}
 
 \label{tab:spec-params}
\end{table}

Starting with these settings, we then considered different variations in frequency and time resolutions (smaller number of mel bands and larger hop sizes). 
Table~\ref{tab:spec-params} shows all different spectrogram configurations that we evaluated on the MTAT dataset. Each configuration results in a different dimension of the resulting feature matrix (the number of mel bands $\times$ the number of frames). 
An audio segment of 29.1 seconds corresponds to
1366 and 1820 frames in the case of no temporal reduction ($\times1$) and the 12 KHz and 16kHz sample rate, respectively.
In turn, the maximum reduction we considered ($\times10$) results in 137 and 182 frames.

All spectrograms were computed using Essentia\footnote{\url{https://essentia.upf.edu}} music audio
analysis library~\cite{bogdanov2013essentia}. It was configured to reproduce 
mel-spectrograms from another analysis library used by the state of the art, LibROSA,\footnote{\url{https://librosa.github.io}} for compatibility.
As a matter of interest, to have a better understanding of what information these spectrograms are able to capture, we provide a number of examples sonifying the resulting mel-spectrograms for all considered frequency and time resolutions online.\footnote{\url{https://andrebola.github.io/EUSIPCO2020/demos}}




\section{Baseline architecture adjustments}\label{sec:model_adjustments}

In this section we explain the changes introduced to the original model architectures presented in Section~\ref{sec:baselines}. 
\subsection{\choi{}}
We try to preserve the original architecture defined in~\cite{choi2016automatic} in terms of the size and number of filters in each layer, but we need to adjust max pooling settings since we are reducing the dimensions of the mel-spectrogram input. We report all such modifications 
for the \choi{} architecture in Table~\ref{tab:vgg_mods}. It reports the sizes of square max-pooling windows in each layer selected accordingly to the number of mel bands and the hop size. We 
prioritize changes in max pooling in the latter layers when possible. We adjust the pooling size to match the input dimensions when possible, 
otherwise padding is applied.
In the case of 16 KHz sample rate, more adjustments to \choi{} are necessary because, having a fixed reference hop size of 256 samples, the higher sample rate implies better temporal resolution and the larger mel-spectrograms (1820 frames).

It is important to note that if we change the resolution of the input, the 3$\times$3 filters in \choi{} capture different ranges of frequency and temporal information. 
For example, they cover twice the mel-frequency range and a doubled time interval when using 48 mel bands and $\times$2 hop size. This can be an advantage, because it reduces the amount of information that the network needs to learn.

\begin{table}[!h]
\caption{Adjusted sizes for max-pooling windows (time and frequency) in the four consecutive layers of the VGG CNN model with respect to the hop size, sample rate and the number of mel bands. The original sizes are highlighted in bold.}
\centering
\begin{tabular}{|c|c|c|}
\hline
\textbf{hop size} & \multicolumn{2}{c|}{\textbf{max-pooling size (time)}} \\
\hline
&\textbf{\textit{12 KHz}} & \textbf{\textit{16 KHz}}\\ 
\hline
$\times$1 & \textbf{4, 5, 8, 8}& 4, 5, 9, 10 \\
$\times$2 & 4, 5, 8, 4            & 4, 5, 9, 5 \\
$\times$3 & 4, 5, 8, 2            &  4, 5, 9, 3 \\
$\times$4 & 4, 5, 8, 2            &  4, 5, 9, 2 \\
$\times$5 &  4, 5, 8, 1           &  4, 5, 9, 2 \\
$\times$10 & 4, 5, 4, 1         &  4, 5, 9, 1\\
& & \\
\hline
 \textbf{\# mel} & \multicolumn{2}{c|}{\textbf{max-pooling size (frequency)}} \\
\hline
128 & \multicolumn{2}{c|}{2, 4, 4, 4} \\
96 & \multicolumn{2}{c|}{\textbf{2, 4, 3, 4}} \\
48 & \multicolumn{2}{c|}{2, 4, 3, 2} \\
32 & \multicolumn{2}{c|}{2, 2, 3, 2} \\
24 & \multicolumn{2}{c|}{2, 2, 3, 2} \\
16 & \multicolumn{2}{c|}{2, 2, 2, 2} \\
8 & \multicolumn{2}{c|}{2, 2, 2, 1} \\
\hline
\end{tabular}
\label{tab:vgg_mods}
\end{table}

\subsection{\pons{}}
In the original model, timbre filters' sizes in frequency are computed relative to the number of mel bands (90\% and 40\%). We preserve the same relation when we change this number. In our implementations we modified the segment size to~3 seconds, as we obtained slightly better results in our preliminary evaluation.\footnote{
Similar to suggestions by other researchers reproducing this model.} 
We keep the temporal dimension of the filters (the number of frames) intact for all considered mel-spectrograms settings.

%
\section{Evaluation Metrics}\label{sec:metrics}
CNN models for auto-tagging output continuous activation values within $[0, 1]$ for each tag, and therefore we can study the performance of binary classifications under different activation thresholds. To this end, following previous works~\cite{oramas2017multi, pons2017end, choi2016automatic} we use Receiver Operating Characteristic Area Under Curve (ROC AUC) averaged across tags as our performance metric. We also report Precision-Recall Area Under Curve (PR AUC), because previous studies~\cite{davis2006relationship} have shown that ROC AUC can give over-optimistic scores when the data is unbalanced, which is our case. Both ROC AUC and PR AUC 
are single value measures characterizing the overall performance, which allows to easily compare multiple systems.

To measure the computational cost of models' training and inference we use an estimate of the number of multiply-accumulate operations required by a network to process one batch (1 GMAC is equal to 1 Giga MAC operations). This metric is related to the time a model requires for training and inference. We use an online tool\footnote{\url{https://dgschwend.github.io/netscope/quickstart.html}} to compute approximate MAC values for our architectures.

\section{Results}\label{sec:results}
\begin{figure*}[htb]
  \centering
  \centerline{\includegraphics[width=\textwidth]{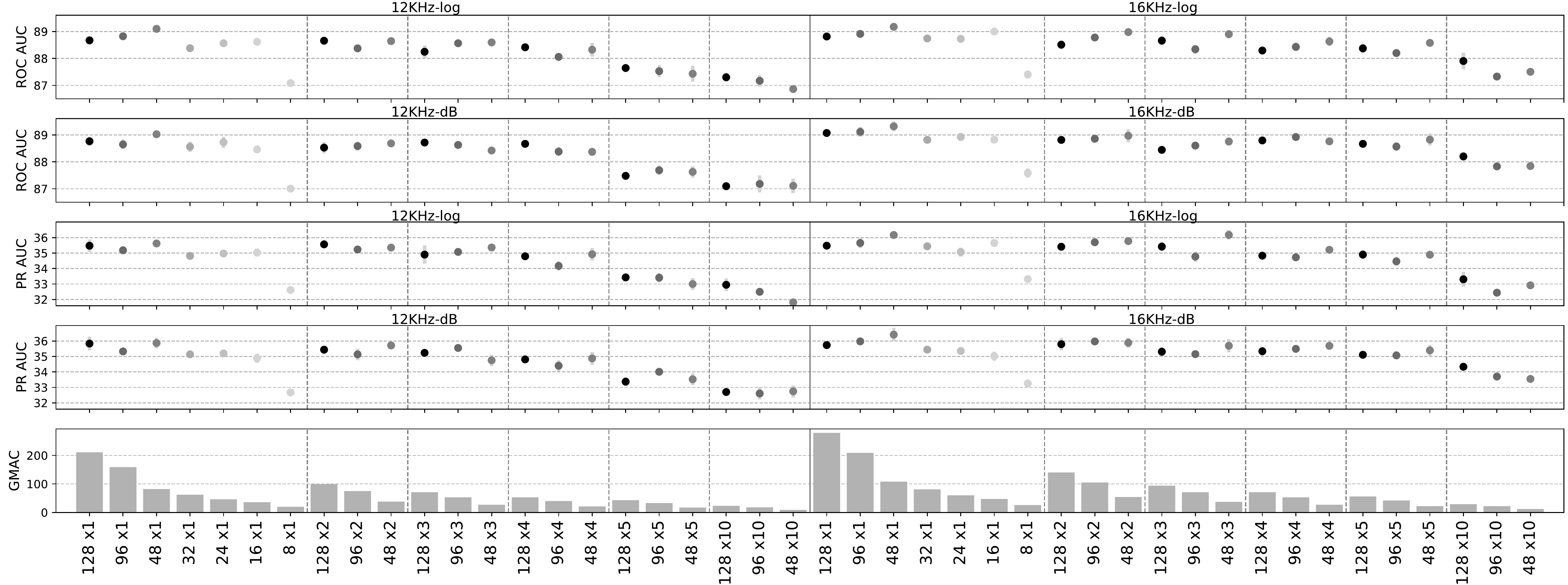}}
%
\caption{Mean and standard deviation of ROC AUC and PR AUC of the \choi{} model computed on three runs for each mel-spectrogram configuration (\# mel, hop size, sample rate, and log type) and the associated GMAC values.}
\label{fig:results}
\end{figure*}

We evaluated the considered mel-spectrogram settings on the adjusted CNN models. Full results for all evaluated configurations are available online.\footnote{\url{https://andrebola.github.io/EUSIPCO2020/results}} In Figure \ref{fig:results} we show the results of the evaluation for \choi{} on the MTAT dataset, repeated three times for each configuration.  The first two plots show the ROC AUC results for the 12 KHz and 16 KHz sample rate using the log and dB scaling. Similarly, the third and forth plots show the PR ROC results under the same conditions. The last plot shows GMAC. 

The results show that using some of the settings we can reduce the size of the input in frequency and time without affecting much the performance of \choi{} on the MTAT dataset. 
For example, if we reduce the frequency resolution from 96 to 48 mel bands we can reduce the MAC operations near 50\% without affecting the performance in all configurations. Similarly, we can also reduce time resolution by 50\% without affecting performance, and in this case we also reduce the MAC operations by 50\% in all configurations.
We can further reduce the number of operations by the cost of some performance decrease. This can be especially useful for applications requiring lightweight models, as we can get a model $\times$10 faster by sacrificing between 1.4 and 1,8\% of the performance depending on the configuration.
Interestingly enough, both ROC AUC and PR AUC slightly improve when using 48 mel bands compared to 96 bands in most of the cases, however no statistically significant difference was found ($P > 0.08$ for all corresponding configurations in an independent samples t-test).

For the 
\pons{} model, we have tested some of the configurations reported in Table~\ref{tab:jordi-res}. We only considered the frequency resolution reduction to 48 mel bands and no hop size increments due to significantly slower training time (see Section~\ref{sec:baselines}). The results show comparable performance of 96- and 48-band mel-spectrograms and are consistent with the above mentioned findings for the \choi{} model. Overall, using 128 mel bands resolution provided the best performance. Also, according to the results, the \pons{} architecture outperforms \choi{}, which is consistent with the reports from the authors.
 
To check how our findings scale, we selected a number of configurations and re-evaluated the models on the MSD dataset. The results are reported in Table~\ref{tab:msd-res}. In the case of \choi{} 
the performance of the baseline architectures is slightly superior to the ones working with lower-resolution mel-spectrograms, which comes by cost of a significantly larger computational effort.
For example, for the 12 KHz sample rate, $\times$ 1 hop size and \textit{dB} compression settings, reducing the number of mel bands from 96 to 48 results in the decrease is 0.16\% in the ROC AUC performance and 50\% reduction in GMACs.
For a similar 16 KHz/\textit{dB} case the reduced model has the same performance with the benefit of twice as low computational speed. 
In the case of the \pons{} architecture we see a reduction of the performance of 0.19\% if we compare 96 vs 48 mel bands using 12 KHz sample rate and 0.11\% for 16 KHz.



\begin{table}[!h]
\caption{ROC AUC and PR AUC of the \pons{} model on the MTAT dataset for a selection of configurations using \textit{dB} log-compression and the reference hop size ($\times$ 1).}
 \begin{center}
 
 \
 \begin{tabular}{|c|c|c|c|}
  \hline
\textbf{\# mels} & \textbf{sample rate} & \textbf{ROC AUC} & \textbf{PR AUC}\\
  \hline
   128 &12 KHz & 90.40 & \textbf{38.54}\\
   96 &12 KHz & \textbf{90.50} & 37.70\\
 48 &12 KHz & 90.33 & 37.80\\
  \hline
 
    128 &16 KHz & \textbf{90.83} & \textbf{38.92} \\
   96 & 16 KHz & 90.60 & 38.09 \\
  48 &16 KHz & 90.50 & 37.70 \\
  \hline
 \end{tabular}
\end{center}
 
 \label{tab:jordi-res}
\end{table}

\begin{table}[!h]
\caption{ROC AUC and PR AUC of the models on the MSD dataset for a selection of configurations using \textit{dB} log-compression.}
\begin{subtable}{\linewidth}
 
\begin{center}
 \begin{tabular}{|c|c|c|c|c|}
  \hline
\textbf{\# mels} & \textbf{hop size} & \textbf{sample rate} & \textbf{ROC AUC} & \textbf{PR AUC}\\
  \hline
 128 & $\times$1&12 KHz& 86.48 & 27.56\\
 96 & $\times$1&12 KHz& \textbf{86.67} & \textbf{27.70}\\
 48 &$\times$1&12 KHz& 86.53 & 27.27\\
  128 & $\times$2&12 KHz& 86.28 & 27.24\\
 96 &$\times$2&12 KHz& 86.18 & 26.93 \\
 48 &$\times$2&12 KHz& 85.86 & 26.42 \\
 \hline
 128 &$\times$1&16 KHz& \textbf{86.84} & \textbf{28.10}\\
 96 &$\times$1&16 KHz& 86.71 & 28.06\\
 48 &$\times$1&16 KHz& 86.73 & 27.78\\
 128 &$\times$2&16 KHz& 86.34 & 27.06\\
  96 &$\times$2&16 KHz& 86.63 & 27.70 \\
  48 &$\times$2&16 KHz&  86.41 & 26.83 \\
\hline
 \end{tabular}
 \end{center}
 \caption{\choi}
 \label{table:choi-msd}
  \end{subtable}
  
 \vspace{0.5cm}

\begin{subtable}{\linewidth}
 \centering
 \small
 \begin{tabular}{|c|c|c|c|c|}
  \hline
\textbf{\# mels} & \textbf{hop size} & \textbf{sample rate} & \textbf{ROC AUC} & \textbf{PR AUC}\\
  \hline
    128 & $\times$1&12 KHz& 87.10 & 26.97\\
  96 &$\times$1&12 KHz& 87.16 & \textbf{27.10} \\
 48 &$\times$1&12 KHz& 86.99 & 26.66 \\
 128 & $\times$1&16 KHz& \textbf{87.21} & 26.91\\
   96 &$\times$1&16 KHz& \textbf{87.21} & 26.96 \\
48 &$\times$1&16 KHz& 87.10 & 26.64 \\
\hline

\end{tabular}
 \caption{\pons{}}
 \label{table:pons-msd}
  \end{subtable}
 
 \label{tab:msd-res}
\end{table}

\section{Conclusions}\label{sec:conclusions}
In this paper we have studied how different mel-spectrogram representations affect the performance of CNN architectures for music auto-tagging. We have compared the performances of two state-of-the-art models when reducing the mel-spectrogram resolution in terms of amount of frequency bands and frame rates. 
We used the MagnaTagaTune dataset for comprehensive performance comparisons and then compared selected configurations on the larger Million Song Dataset. 
The results suggest that is possible to preserve a similar performance while reducing the size of the input. They 
can 
help researchers and practitioners 
to make trade-off decision between accuracy of the models, data storage size and training and inference time, 
which are crucial in 
many applications.

As a future work, other approaches such as quantization of mel-spectrogram values will be considered for the reduction of the input data dimensionality. The conducted evaluation can be also extended to other state-of-the-art architectures operating on mel-spectrograms~\cite{choi2017convolutional}, constant-Q~\cite{oramas2018multimodal} and raw waveform approaches~\cite{lee2017sample,choi2018comparison}. It is also promising to conduct a similar evaluation on other audio auto-tagging tasks. 
All the code to reproduce this study is open-source and available online.\footnote{\url{https://andrebola.github.io/EUSIPCO2020/}}


\section*{Acknowledgment}
This research has been supported by Kakao Corp.

\bibliographystyle{IEEEbib}
\bibliography{strings,refs}

\end{document}